\documentstyle[12pt]{article}

\font\twlgot =eufm10 scaled \magstep1
\font\egtgot =eufm8
\font\sevgot =eufm7
\font\twlmsb =msbm10 scaled \magstep1
\font\egtmsb =msbm8
\font\sevmsb =msbm7

\newfam\gotfam
\def\pgot{\fam\gotfam\twlgot}
\textfont\gotfam\twlgot
\scriptfont\gotfam\egtgot
\scriptscriptfont\gotfam\sevgot
\def\got{\protect\pgot}
\newfam\msbfam
\textfont\msbfam\twlmsb
\scriptfont\msbfam\egtmsb
\scriptscriptfont\msbfam\sevmsb
\def\Bbb{\protect\pBbb}
\def\pBbb{\relax\ifmmode\expandafter\Bb\else\typeout{You cann't use
Bbb in text mode}\fi}
\def\Bb #1{{\fam\msbfam\relax#1}}

\newcommand{\gO}{{\got O}}
\newcommand{\gd}{{\got d}}
\newcommand{\gQ}{{\got T}}

\def\thebibliography#1{\section*{References}\list
  {[\arabic{enumi}]}{\settowidth\labelwidth{#1}\leftmargin\labelwidth
    \advance\leftmargin\labelsep
    \usecounter{enumi}}
    \def\newblock{\hskip .11em plus .33em minus .07em}
    \sloppy\clubpenalty4000\widowpenalty4000
    \sfcode`\.=1000\relax}

\def\op#1{\mathop{\fam0 #1}\limits}

\newcommand{\id}{{\rm Id\,}}

\newcommand{\nm}[1]{\mid {#1}\mid}
\newcommand{\beq}{\begin{equation}}
\newcommand{\eeq}{\end{equation}}
\newcommand{\ben}{\begin{eqnarray}}
\newcommand{\een}{\end{eqnarray}}
\newcommand{\be}{\begin{eqnarray*}}
\newcommand{\ee}{\end{eqnarray*}}
\newcommand{\bea}{\begin{eqalph}}
\newcommand{\eea}{\end{eqalph}}

\newcommand{\cL}{{\cal L}}

\newcommand{\cO}{{\cal O}}
\newcommand{\cQ}{{\cal Q}}
\newcommand{\bL}{{\bf L}}

\newcommand{\vr}{\varrho}

\newcommand{\dl}{\delta}
\newcommand{\la}{\lambda}
\newcommand{\La}{\Lambda}
\newcommand{\f}{\phi}
\newcommand{\om}{\omega}

\newcommand{\m}{\mu}

\newcommand{\th}{\theta}

\newcommand{\vf}{\varphi}
\newcommand{\up}{\upsilon}

\newcommand{\di}{{\rm dim\,}}

\newcommand{\si}{\sigma}
\newcommand{\Si}{\Sigma}
\newcommand{\w}{\wedge}

\newcommand{\ol}{\overline}
\newcommand{\dr}{\partial}
\newcommand{\ar}{\op\longrightarrow}
\newcommand{\ot}{\otimes}
\newcommand{\ap}{\approx}

\newcounter{eqalph}
\newcounter{equationa}
\newcounter{theorem}
\newcounter{remark}
\newcounter{proposition}
\newcounter{lemma}
\newcounter{corollary}
\newcounter{definition}

\newenvironment{eqalph}{\stepcounter{equation}
\setcounter{equationa}{\value{equation}}
\setcounter{equation}{0}

\begin{eqnarray}}{\end{eqnarray}\setcounter{equation}{\value{equationa}}}

\def\theremark{\arabic{remark}}

\def\thedefinition{\arabic{definition}}

\newcommand{\mar}[1]{}

\begin{document}
\hbox{}

{\parindent=0pt

{\large\bf Generalized Lagrangian symmetries 
depending on higher order \\ derivatives. Conservation laws and the
characteristic equation}
\bigskip

{\bf Giovanni Giachetta}$^1$, {\bf Luigi Mangiarotti}$^1$ {\bf
and Gennadi Sardanashvily}$^2$
\bigskip

\begin{small}
$^1$ Department of Mathematics and Informatics, University of
Camerino, 62032 Camerino (MC), Italy 

$^2$ Department of Theoretical Physics, Physics Faculty, Moscow
State University, 117234 Moscow, Russia
\medskip

E-mail: giovanni.giachetta@unicam.it,
luigi.mangiarotti@unicam.it and sard@grav.phys.msu.su

\end{small}

\bigskip

{\bf Abstract.}

Given a finite order Lagrangian $L$ on a fibre bundle,
its global generalized symmetries depending on higher order
derivatives of dynamic variables are considered. The first
variational formula is obtained. It leads both to the
corresponding Lagrangian conservation laws and the
characteristic equation for generalized symmetries of $L$.  
\medskip



}

\bigskip
\bigskip
 
Symmetries of differential equations under transformations
of dynamic variables depending on their derivatives 
have been intensively investigated (see, e.g.,
\cite{and93,kras,olv} for a survey). Following 
\cite{and93,olv}, we agree to call them the generalized
symmetries. In \cite{olv}, generalized symmetries of Lagrangian
systems and the corresponding conservation laws on a local
coordinate chart are described in detail. The recent work
\cite{fat} turns to the global analysis of first order
Lagrangian systems and conservation laws under generalized
symmetries depending on first order derivatives, but the
symmetry condition is imposed on the Poincar\'e--Cartan form of
a Lagrangian. In analytical mechanics, generalized
symmetries and the corresponding conserved quantities (e.g., the
Runge--Lenz vector in the Kepler problem) are well known
\cite{olv}. In application to field theory, let us mention the
Lie derivative (the Kosmann lift) of Dirac spinor fields
\cite{fat99,book00,sard}, the Poisson sigma model
\cite{fulp}, and BRST transformations \cite{fat,fulp}. The
latter however involve the notion of jets of functions on graded
manifolds \cite{book00,mpl} which is beyond the scope of this
work. Our goal here is to study the conservation laws in higher
order  Lagrangian systems on fibre bundles under generalized 
symmetries depending on derivatives of any finite order.
 
There are different approaches to the study of
Lagrangian conservation laws. We are based on the first
variational formula (see \cite{book,book00,epr} for a survey).

Let $Y\to X$ be a smooth fibre bundle coordinated by
$(x^\la,y^i)$.  In Lagrangian formalism on $Y\to X$, an
$r$-order Lagrangian is defined as a horizontal density
\mar{g10'}\beq
L=\cL\om, \qquad \om=dx^1\w\cdots\w dx^n, \qquad n=\di X,
\label{g10'}
\eeq
on the $r$-order jet manifold $J^rX$ of sections of $Y\to X$.
This manifold is equipped with the adapted coordinates
$(x^\la, y^i,y^i_\La)$, $0<\nm\La \leq r$, where a
multi-index
$\La$,
$\nm\La=k$, denotes a collection of numbers $(\la_k...\la_1)$
modulo permutations. By $\La+\Si$ is meant the
collection $(\la_k\cdots\la_1\si_s\cdots\si_1)$ 
modulo permutations. We use the notation 
\mar{5.177}\beq
d_\la = \dr_\la + \op\sum_{|\La|\geq 0}
y^i_{\la+\La}\dr_i^\La, \qquad
d_\La=d_{\la_r}\circ\cdots\circ d_{\la_1}, \quad
\La=(\la_r...\la_1). \label{5.177}
\eeq

In order to obtain Noether conservation laws, one 
considers vector fields 
\mar{g15}\beq
u=u^\la(x^\m)\dr_\la +u^i(x^\m,y^j)\dr_i \label{g15}
\eeq
on $Y$ projected onto $X$. They are infinitesimal generators of
local one-parameter groups of bundle automorphisms 
of $Y\to X$. Their jet prolongation
onto $J^rY$ read
\mar{55.5}\beq
J^r u =
u^\la\dr_\la + u^i\dr_i +\op\sum_{0<|\La| \leq r}
u_\La^i\dr_i^\La, \qquad u_\La^i = d_\La(u^i - y_\m^i u^\m)
+y^i_{\m+\La}u^\m.
\label{55.5}
\eeq
One says that $u$ (\ref{g15}) is a (variational) symmetry 
of a Lagrangian 
$L$ (\ref{g10'}) if the Lie derivative
$\bL_{J^ru}L$ of $L$ along $J^r u$ vanishes. The first
variational formula provides the canonical decomposition of this
Lie derivative 
\mar{g16}\beq
\bL_{J^ru}L=u_V\rfloor\dl L + d_H(h_0(J^{2r-1}u\rfloor\Xi_L)), 
\label{g16}
\eeq
where $\dl L$ is the Euler--Lagrange operator of $L$, $\Xi_L$ is
a Poincar\'e--Cartan form of $L$, $u_V$ is a vertical part of
$u$,
$d_H$ is the total (horizontal) differential, and $h_0$ is the
horizontal projection (see the definitions below). If $u$ is a
symmetry of $L$, the first variational
formula (\ref{g16}) restricted to the  kernel of
the Euler--Lagrange operator $\dl L$ leads to the Noether
conservation law
\be
0\ap d_H(h_0(J^ru\rfloor\Xi_L)).
\ee
A vector field $u$ (\ref{g15}) is said
to be a divergence symmetry of $L$ if the Lie
derivative 
$\bL_{J^ru}L$ is a total differential $d_H\si$. Then, the
first variational formula (\ref{g16}) on Ker$\,\dl L$ provides
the generalized Noether conservation law
\mar{g19}\beq
0\ap d_H(h_0(J^ru\rfloor\Xi_L)-\si). \label{g19}
\eeq
Note that any divergence symmetry of $L$ is a symmetry of the
Euler--Lagrange operator (i.e., $\bL_{J^{2r}u}\dl L=0$) as 
it follows from  the master identity
\mar{g44}\beq
\dl(\bL_{J^ru}L)= \bL_{J^{2r}u}\dl L \label{g44}
\eeq 
and the equality 
\mar{g20}\beq
\dl(\bL_{J^ru}L)=\dl(d_H\si)=0. \label{g20}
\eeq

We aim at extending the first variational
formula (\ref{g16}) to the above mentioned generalized
symmetries (see the formula (\ref{g8}) below) and obtaining the
corresponding conservation laws (\ref{g32}). Herewith, the
equality (\ref{g33}), similar to (\ref{g20}), 
gives an equation for divergence symmetries of a
Lagrangian $L$. 

It should be emphasized the
following.

(i) Infinitesimal generalized symmetries, called
generalized vector fields, take the local coordinate form 
\mar{g21}\beq
u=u^\la(x^\m,y^j,y^j_\La)\dr_\la +u^i(x^\m,y^j,y^j_\La)\dr_i,
\label{g21}
\eeq 
where $u^\la$, $u^i$ are local functions on some
finite order jet manifold
$J^kY$. Their $r$-order jet prolongation $J^ru$ is given by the
formula (\ref{55.5}). However, the generalized vector fields
(\ref{g21}) fail to be vector fields on a finite order jet
manifold. Namely, $J^ru$  is a derivation of the ring
$C^\infty(J^rY$) of smooth real functions on
the jet manifold
$J^rY$ which takes its values in the ring of smooth real
functions on  the $(r+k)$-order jet manifold $J^{r+k}Y$. In
\cite{olv}, generalized vector fields (\ref{g21}) are 
locally introduced as formal differential operators. In
\cite{fat}, they are associated to sections of the pull-back
bundle
$TY\times_Y J^kY\to J^kY$.  We describe generalized symmetries
in the framework of infinite order jet formalism.  

(ii) The invariance of a Lagrangian under the transformations
(\ref{g21}) imposes rather strong conditions on these
transformations. Therefore, one allows generalized
symmetries to be the divergence symmetries of a Lagrangian.

Infinite order jet formalism provides a
convenient tool for studying Lagrangian systems of
unspecified finite order \cite{book,kras,book00,epr,sau}. 

Finite order jet manifolds make up the 
inverse system
\mar{5.10}\beq
X\op\longleftarrow^\pi Y\op\longleftarrow^{\pi^1_0} J^1Y\cdots
\longleftarrow J^{r-1}Y \op\longleftarrow^{\pi^r_{r-1}}
J^rY\longleftarrow\cdots.
\label{5.10}
\eeq
Its projective limit $J^\infty Y$,
called the infinite order jet space,
is endowed with the weakest
topology such that surjections $\pi^\infty_r:J^\infty Y\to
J^rY$ are continuous. This
topology makes $J^\infty Y$ into a paracompact Fr\'echet
manifold
\cite{tak2}. A bundle coordinate atlas
$\{U_Y,(x^\la,y^i)\}$ of $Y\to X$ yields the manifold
coordinate atlas
\mar{jet1}\beq
\{(\pi^\infty_0)^{-1}(U_Y), (x^\la, y^i_\La)\}, \qquad
{y'}^i_{\la+\La}=\frac{\dr x^\m}{\dr x'^\la}d_\m y'^i_\La,
\qquad
0\leq|\La|,
\label{jet1}
\eeq
of $J^\infty Y$, where
$d_\la$ are the total derivatives (\ref{5.177}).

Let
$\cO_r^*$ denote the graded differential algebra of exterior
forms on the jet manifold $J^rY$. With the inverse system
(\ref{5.10}), we have the
direct system of
$C^\infty(X)$-modules
\mar{5.7}\beq
\cO^*(X)\op\longrightarrow^{\pi^*} \cO^*(Y) 
\op\longrightarrow^{\pi^1_0{}^*} \cO_1^* \cdots
\op\longrightarrow^{\pi^r_{r-1}{}^*}
 \cO_r^* \longrightarrow\cdots \label{5.7}
\eeq
where $\pi^{r+1}_r{}^*$ are the pull-back monomorphisms.
Its direct limit $\cO_\infty^*$ is a graded differential
algebra, whose $d$-cohomology is proved to be
isomorphic to
the de Rham cohomology $H^*(Y)$ of a fibre bundle $Y$
\cite{ander}.

Though $J^\infty Y$ is not a smooth manifold, 
one can think of elements of $\cO_\infty^*$
as being objects on $J^\infty Y$ as
follows. Let $\gO^*_r$ be the sheaf
of germs of exterior forms on the $r$-order jet 
manifold $J^rY$, and let
$\ol\gO^*_r$ be its canonical presheaf.  There is the direct 
system of presheaves
\be
\ol\gO^*_X\op\longrightarrow^{\pi^*} \ol\gO^*_0 
\op\longrightarrow^{\pi^1_0{}^*} \ol\gO_1^* \cdots
\op\longrightarrow^{\pi^r_{r-1}{}^*}
 \ol\gO_r^* \longrightarrow\cdots. 
\ee
Its direct limit $\ol\gO^*_\infty$ 
is a presheaf of graded differential
algebras on
$J^\infty Y$. Let $\gQ^*_\infty$ be a sheaf constructed from 
$\ol\gO^*_\infty$. The algebra 
$\cQ^*_\infty$ of 
sections of $\gQ^*_\infty$ is a graded differential
algebra whose elements
$\f$ possess the following property.
For any point $q\in J^\infty Y$, there exist an open
neighbourhood $U$ of $q$ and an
exterior form
$\f^{(k)}$ on some finite order jet manifold $J^kY$ such that
$\f|_U= \f^{(k)}\circ \pi^\infty_k|_U$. We  agree to call
elements of  $\cQ^*_\infty$ the exterior forms of locally
finite jet order on $J^\infty Y$. There is the natural
monomorphism
$\cO^*_\infty
\to\cQ^*_\infty$ whose image consists of the pull-back onto
$J^\infty Y$ of exterior forms on finite order jet manifolds.

Restricted to a
coordinate chart (\ref{jet1}), elements of
$\cO^*_\infty$ can be written in a
coordinate form, where horizontal forms 
$\{dx^\la\}$ and contact 1-forms
$\{\th^i_\La=dy^i_\La -y^i_{\la+\La}dx^\la\}$ make up local
generators of the algebra
$\cO^*_\infty$. 
There is the canonical decomposition
\be
\cO^*_\infty =\op\oplus_{k,s}\cO^{k,s}_\infty, \qquad 0\leq k,
\qquad 0\leq s\leq n,
\ee
of $\cO^*_\infty$ into $\cO^0_\infty$-modules $\cO^{k,s}_\infty$
of $k$-contact and $s$-horizontal forms
together with the corresponding
projections
\be
h_k:\cO^*_\infty\to \cO^{k,*}_\infty, \quad 0\leq k, \qquad
h^s:\cO^*_\infty\to \cO^{*,s}_\infty, \quad 0\leq s
\leq n.
\ee 
Accordingly, the
exterior differential on $\cO_\infty^*$ is split
into the sum $d=d_H+d_V$ of horizontal and vertical
differentials 
\be
&& d_H\circ h_k=h_k\circ d\circ h_k, \qquad d_H(\f)=
dx^\la\w d_\la(\f), \\ 
&& d_V \circ h^s=h^s\circ d\circ h^s, \qquad
d_V(\f)=\th^i_\La \w \dr^\La_i\f, \qquad \f\in\cO^*_\infty.
\ee
In particular, we have the
relations
\be
d_H\circ h_0=h_0\circ d,\qquad
 d_H(dx^\m) =0, \qquad 
d_H(\th^i_\La)= dx^\la\w \th^i_{\la+\La}.
\ee

Furthermore, one defines the projection $\Bbb R$-module
endomorphism 
\mar{r12}\beq
\vr=\op\sum_{k>0} \frac1k\ol\vr\circ h_k\circ h^n,
\qquad \ol\vr(\f)= \op\sum_{|\La|\geq 0}
(-1)^{\nm\La}\th^i\w [d_\La(\dr^\La_i\rfloor\f)], 
\qquad \f\in \cO^{>0,n}_\infty, \label{r12}
\eeq
of $\cO^*_\infty$ such that
$\vr\circ d_H=0$ and $\vr\circ d\circ \vr -\vr\circ
d=0$ (e.g., \cite{bau,book,tul}). Put
$E_k=\vr(\cO^{k,n}_\infty)$, $k>0$.
Then, the variational operator on $\cO^{*,n}_\infty$ is defined
as the morphism $\dl=\vr\circ d$. 
It is nilpotent, and obeys the relation 
$\dl\circ\vr-\vr\circ d=0$. As a consequence, 
the graded differential algebra
$\cO^*_\infty$ is split into the so called variational
bicomplex. Here, we are concerned only with its subcomplexes
\mar{b317,xx}\ben
&& 0\to\Bbb R\to \cO^0_\infty
\ar^{d_H}\cO^{0,1}_\infty\ar^{d_H}\cdots  
\op\longrightarrow^{d_H} 
\cO^{0,n}_\infty  \op\longrightarrow^\dl E_1 
\op\longrightarrow^\dl 
E_2 \longrightarrow \cdots,  \label{b317} \\
&& 0\to \cO^{1,0}_\infty\ar^{d_H} \cO^{1,1}_\infty
\ar^{d_H}\cdots
\ar^{d_H}\cO^{1,n}_\infty\ar^\vr E_1\to 0.
\label{xx}
\een
The first one, called the variational complex,
provides the algebraic approach to the calculus of variations
in the class of exterior forms of finite jet order. Namely,
one can think of an element $L\in \cO^{0,n}_\infty$ as being
a finite order Lagrangian, while the
variational operator $\dl$ acting on $\cO^{0,n}_\infty$ is the
Euler--Lagrange operator
\mar{g1}\beq
\dl L=\dl_i\cL \th^i\w\om =\op\sum_{|\La|\geq
0}(-1)^{|\La|}d_\La\dr^\La_i \cL\th^i\w\om. \label{g1}
\eeq

The key point is that the paracompact space
$J^\infty Y$ admits a partition of unity by functions 
$f\in\cQ^0_\infty$ of locally finite jet order \cite{tak2}. It
follows that the sheaves of
$\cQ^0_\infty$-modules on
$J^\infty Y$ are fine and acyclic. Therefore, the 
abstract de Rham theorem on cohomology of a sheaf resolution 
enables one to obtain the cohomology of the
graded differential algebra $\cQ^*_\infty$
\cite{and,ander,tak2}. Furthermore, the
$d$-,
$d_H$- and $\dl$-cohomology of
its subalgebra  $\cO^*_\infty$ 
is proved to coincide with that of $\cQ^*_\infty$ 
\cite{jmp,ijmms,epr}.
As a consequence, one can show that cohomology of the
variational complex  (\ref{b317}) is isomorphic  
to the de Rham cohomology of a fibre bundle
$Y$, i.e.,
\be
H^{k<n}(d_H;\cO^*_\infty)=H^{k<n}(Y), \qquad H^{k-n}(\dl;
\cO^*_\infty)=H^{k\geq n}(Y)
\ee
\cite{and,jmp,ijmms,epr}. 
It follows that, in particular, 
any $\dl$-closed form $L\in\cO^{0,n}$ is split into the sum
\mar{t42}\beq
 L=h_0\varphi + d_H\xi,  \qquad \xi\in
\cO^{0,n-1}_\infty,
\label{t42}
\eeq
where $\varphi$ is a closed $n$-form on $Y$. In other words, 
a finite order Lagrangian
$L$  is variationally trivial iff
it takes the form (\ref{t42}).

Similarly, one can show that the complex (\ref{xx}) is
exact  
\cite{jmp,ijmms,epr,tak2}. Its exactness at the term
$\cO^{1,n}_\infty$ implies that, if $\vr(\f)=0$, $\f\in
\cO^{1,n}_\infty$, then
$\f=d_H\xi$, $\xi\in \cO^{1,n-1}_\infty$. Since $\vr$ is a
projection operator, 
there is the $\Bbb R$-module decomposition
\mar{30jpa}\beq
\cO^{1,n}_\infty=E_1\oplus d_H(\cO^{1,n-1}_\infty).
\label{30jpa}
\eeq
Given a Lagrangian $L\in \cO^{0,n}_\infty$, the decomposition
(\ref{30jpa}) 
provides its splitting
\mar{+421}\beq
dL=\vr(dL) + (\id -\vr)(dL)= \dl L - d_H(\Xi), \label{+421}
\eeq
where $\Xi\in \cO^{1,n-1}_\infty$. This splitting leads to the
first variational formula as follows.

Let us consider derivations $\up\in\gd\cO^0_\infty$ of the ring
$\cO^0_\infty$ of smooth functions of finite jet order on
$J^\infty Y$. With respect to the atlas
(\ref{jet1}), they are given by the coordinate expression
\mar{g3}\beq
\up=\up^\la \dr_\la + \up^i\dr_i +
\op\sum_{|\La|>0}\up^i_\La
\dr^\La_i, \label{g3}  
\eeq
where components $\up^\la$, $\up^i$, $\up^i_\La$ are
local smooth functions of finite jet order on $J^\infty Y$
which obey the transformation law
\be
\up'^\la=\frac{\dr x'^\la}{\dr x^\m}\up^\m, \qquad
\up'^i=\frac{\dr y'^i}{\dr y^j}\up^j + \frac{\dr y'^i}{\dr
x^\m}\up^\m, \qquad 
\up'^i_\La=\op\sum_{|\Si|\leq|\La|}\frac{\dr y'^i_\La}{\dr
y^j_\Si}\up^j_\Si +
\frac{\dr y'^i_\La}{\dr x^\m}\up^\m. 
\ee
The interior product $\up\rfloor\f$, 
and the Lie derivative $\bL_\up\f$,
$\f\in\cO^*_\infty$ are defined in a
standard way. 
An element $\up\in\gd\cO^0_\infty$ is said to be a generalized
symmetry if the Lie derivative of any contact one-form
$\th\in \cO^{1,0}$ along
$\up$ again is a contact form. One can easily justify that
$\up\in\gd\cO^0_\infty$ is a generalized symmetry iff it is
given by the coordinate expression (\ref{g3}) where
\mar{g4}\beq
\up^i_\La=d_\La(\up^i-y^i_\m\up^\m)+y^i_{\m+\La}\up^\m, \qquad
0<|\La|.
\label{g4}
\eeq
For instance,  let $\tau$ be a vector field on $X$. Then,
the derivation 
$\tau\rfloor (d_H f)$, $f\in \cO_\infty^0$,
of the ring $\cO_\infty^0$ is a generalized symmetry
\mar{g2}\beq
J^\infty\tau=\tau^\m d_\m. \label{g2}
\eeq
Moreover, any generalized symmetry $\up$ is
brought into the form
\mar{g5}\beq
\up=\up_H +\up_V=\up^\la d_\la + (\ol\up^i\dr_i +
\sum\ol\up^i_\La\dr^\La_i), \qquad \ol\up^i=
\up^i-y^i_\m\up^\m, \qquad \ol\up^i_\La=d_\La\ol\up^i.
\label{g5}
\eeq
This is the horizontal splitting of 
$\up$ with respect to the canonical connection
$\nabla=dx^\la\ot d_\la$ on the $C^\infty(X)$-ring
$\cO^0_\infty$ \cite{book00}.
In particular, if $\up^\la=0$, we have the
relation
\mar{g6}\beq
\up\rfloor d_H\f=-d_H(\up\rfloor\f), \qquad \f\in\cO^*_\infty.
\label{g6}
\eeq 

Let us 
consider the Lie derivative
\mar{g7}\beq 
\bL_\up L=\up\rfloor dL + d(\up\rfloor L)
=\up_V\rfloor dL + d_H(\up_H\rfloor L) +\cL d_V
(\up_H\rfloor\om)
\label{g7} 
\eeq
of a Lagrangian $L\in\cO^{0,n}_\infty$ along a generalized
symmetry
$\up$ (\ref{g5}). Using the splitting (\ref{+421})
and the equality (\ref{g6}), we come to the desired
first variational formula
\mar{g8}\ben
&& \bL_\up L=
 \up_V\rfloor\dl L -\up_V\rfloor
d_H\Xi + d_H(\up_H\rfloor L) +\cL d_V (\up_H\rfloor\om)
= \nonumber\\
&& \qquad \up_V\rfloor\dl L +d_H(\up_V\rfloor\Xi +
\up_H\rfloor L) +\cL d_V (\up_H\rfloor\om)=\nonumber\\
&& \qquad \up_V\rfloor\dl L +d_H(h_0(\up\rfloor\Xi_L)) +\cL d_V
(\up_H\rfloor\om), 
\label{g8}
\een
where $\Xi_L$ is some Poincar\'e--Cartan form of a finite order
Lagrangian $L$.

Let a generalized symmetry $\up$ (\ref{g5}) be a divergence
symmetry of a Lagrangian
$L$, i.e., 
\mar{g30}\beq
\bL_\up L=d_H\si, \qquad \si\in \cO^{0,n-1}_\infty.
\label{g30}
\eeq
By virtue of the expression (\ref{g7}), this condition implies
that a generalized symmetry $\up$ is projected onto $X$, i.e., 
its components $\up^\la$ depend only on coordinates on $X$. 
Then, the first variational formula (\ref{g8}) takes the form 
\mar{g11}\beq
d_H\si= \up_V\rfloor\dl L +d_H(h_0(\up\rfloor\Xi_L)).
\label{g11}
\eeq
Restricted to Ker$\,\dl L$, it leads to the generalized Noether
conservation law
\mar{g32}\beq
0\ap d_H(h_0(\up\rfloor\Xi_L)-\si). \label{g32}
\eeq

 A glance at the expression (\ref{g7}) shows
that a generalized symmetry $\up$ (\ref{g5}) projected onto $X$
is a divergence symmetry of a Lagrangian $L$ iff its vertical
part $\up_V$ is so. Moreover, $\up$ and $\up_V$ lead to
the same  generalized Noether conservation law (\ref{g32}). For
instance, if
$\up=J^\infty\tau$ (\ref{g2}), the first variational formula
(\ref{g8}) and the conservation law (\ref{g32}) become
tautological.

Note that a Poincar\'e--Cartan form $\Xi_L$ of an $r$-order
Lagrangian
$L=\cL\om$ fails to be uniquely defined unless $r=1$ or $\di
X=1$. It is given by the coordinate expression
\mar{g43}\ben
&& \Xi_L=\cL\om
+\op\sum^{r-1}_{s=0}F^{\la\m_s\ldots\m_1}_i
\th^i_{\m_s\ldots\m_1}\w\om_\la, \qquad
\om_\la=\dr_\la\rfloor\om, \label{g43}\\
&& F_i^{\nu_r\ldots\nu_1}=\dr_i^{\nu_r\ldots\nu_1}\cL, \qquad  
F_i^{\nu_k\ldots\nu_1}=
\dr_i^{\nu_k\ldots\nu_1}\cL-d_\la F_i^{\la\nu_k\ldots\nu_1}
+c_i^{\nu_k\ldots\nu_1}, \quad 1\leq k<r, \nonumber
\een
where the functions $c_i^{\nu_k\ldots\nu_1}$, of jet order at
most $2r-k-1$, satisfy the condition
$c_i^{(\nu_k\nu_{k-1})\ldots\nu_1}=0$ and $c^\nu_i=0$
\cite{got,krupk}. Any
Poincar\'e--Cartan form (\ref{g43}) can be locally brought into
the form where all the functions $c_i^{\ldots}$ equal zero.
This local expression is used in \cite{olv}. It is globally
valid if either $L$ is of first order (see \cite{fat}) or
$\di X=1$, i.e., in the higher order mechanics.

One can obtain the characteristic equation for
divergence symmetries of a Lagrangian $L$ as follows. Let a
generalized symmetry $\up$ (\ref{g5}) be projected onto $X$.
Then, the Lie derivative $\bL_\up L$ (\ref{g7}) is a horizontal
density. Let us require that it is a $\dl$-closed form, i.e., 
\mar{g33}\beq
\dl(\bL_\up L)=0. \label{g33}
\eeq
In accordance with the equality (\ref{t42}), this condition is
fulfilled iff 
\be
\bL_\up L=h_0\vf +d_H\si,
\ee
where $\vf$ is a closed form on $Y$. It follows that $\up$
is a divergence symmetry of
$L$ at least locally. Thus, the equation (\ref{g33}) enables
one to find all divergence symmetries of a given Lagrangian
$L$. Note that the master identity (\ref{g44}) fails to be true
for generalized symmetries. There is the local relation
\be
\dl(\bL_\up L)=\bL_\up\dl L + \op\sum_{|\La|>0}(-1)^{|\La|}d_\La
(\dr^\La_k\up^i\dl_i\cL dy^k)\w\om 
\ee
used in \cite{olv}.

\end{document}